# Construction of stock molecular system and popularization of Density Functional Theory in stock market

Huajian Li[a, *], Longjian Li[b], Jiajian Liang[c]

[a] School of Chemistry and Chemical Engineering, Guangzhou University, Guangzhou 510006, P. R. China
[b] School of Civil and Transportation Engineering, Guangdong University of Technology, Guangzhou 510006, P. R. China
[c] School of Mechanical and Electrical Engineering, Guangzhou University, Guangzhou 510006, P. R. China

**ABSTRACT**

Over the past two decades, some scholars have noticed the correlation between quantum mechanics and finance/economy, making some novel attempts to introduce the theoretical framework of quantum mechanics into financial and economic research, subsequently a new research domain called quantum finance or quantum economy was set up. In particular, some studies have made their endeavour in the stock market, utilizing the quantum mechanical paradigm to describe the movement of stock price. Nevertheless, the majority of researches have paid attention to describing the motion of a single stock, and drawn an analogy between the motion of a single stock and a one-dimensional infinite well, or one-dimensional harmonic oscillator model, whose modality looks alike to the one-electron Schrödinger equation, in which the information is solved analytically in most cases. Hitherto, the whole stock market system composed of all stocks and stock indexes have not been discussed. In this paper, the concept of stock molecular system is first proposed with pioneer. The modality of stock molecular system resembles the multi-electrons Schrödinger equation with Born-Oppenheimer approximation. Similar to the interaction among all nuclei and electrons in a molecule, the interaction exist among all stock indexes and stocks. This paper also establish the stock-index Coulomb potential, stock-index Coulomb potential, stock-stock Coulomb potential and stock coulomb correlation terms by statistical theory. At length, the conceive and feasibility of drawing upon density functional theory (DFT) to solve the Schrödinger equation of stock molecular system are put forward together with proof, ending up with experiments executed in CSI 300 index system.

## 1. INTRODUCTION

Theories in the domain of physics were applied to the study of economy and finance, probably inaugurating in the 1990s. To start with, Mantegna [1,2], Durlauf [3] and Ilinski [4] etc. successively found that there is strong correlation between statistical physics theories and economic/financial phenomena to some extent, thereby making an attempt to apply some statistical physics models into economic/financial analysis. Simultaneously, other scholars also sought the assistance from other branches of physics. Since the economic/financial phenomena are highly associated with the movements of microscopic particles, which behave randomly. Moreover, there exists interaction among each other of economic/financial objects (similar to quantum effects), for the reason theories of quantum mechanics are a suitable tool to study economy and finance. Dash [5] systematically proposed the application of Feynman path integral [6] in option pricing, followed by a group of scholars emerging to explore the application of path integral tools in economy and finance [7,8]. Baaquie [8] proposed the Schrödinger equation form of Black-Scholes option pricing formula [9] with its Hamiltonian and Lagrangian. Haven [10] embedded Black-Scholes option pricing formula into the framework of quantum mechanics and put forward the arbitrage potential function. Schaden [11] conducted the quantitative analysis of financial markets under the framework of quantum mechanics, taking a series of wave functions representing all possible investment portfolios of investors as the basis of Hilbert space. Zhang et al. [12] employed the one-dimensional infinite well model for the portrayal of single-stock price motion, wherein a cosine periodic potential function was introduced into the Hamiltonian which symbolized the external information affecting the stock price, such as national policies, natural disasters, business cycle and investor psychology etc. Cotfas [13] developed the quantum model of Zhang [12], taking the stock return instead of the stock price as variable of the wave function, and discretized the stock return to make the stock system with quantized properties like a particle system. Ye et al. [14] brought up a non-classical one-dimensional harmonic oscillator model to describe the fluctuation of stock price, meanwhile pointing out that the probability density function of the system is Gaussian distribution when in the ground state. Chukiat et al. [15] Applied quantum mechanics theories to analyze the risk and return of the Association of Southeast Asian Nations (ASEAN) stock market, as well comparing it with traditional econometrics. K.hn et al. [16] wielded the harmonic oscillator quantum model to depict the fluctuation of the FTSE All Share Index, eventually concluded with smaller fitting errors and better goodness-of-fit than traditional models of stochastic process (geometric Brownian motion model and Heston model). Different from previous harmonic oscillator quantum model, Gao [17] proposed a non-harmonic oscillator model to portray the real stock behavior, inside the Hamiltonian a polynomial potential energy term which is a function of stock return was affiliated, with a explainable financial significance.

* Corresponding author.
 *E-mail address:* 2112005065@e.gzhu.edu.cn (Huajian Li)

However, possibly ascribed to the inconvenience of getting hold of an analytical solution, the aforementioned methods merely focus on the motion of single stock, omitting the correlation among all stock indexes and all stocks similar in appearance to the one-electron Schrödinger equation. Hence, we proposes a stock molecular system consisted of all stocks and stock indexes. All stock indexes are regarded as nuclei in a molecule, and the constituent stocks of each stock index are treated as electrons surrounding their respective stock index which resemble the multi-electron Schrödinger equation of a molecular system. In particular, the Born-Oppenheimer approximation is imported into the corresponding Hamiltonian. In this paper, we employed the regression statistical approach to construct the stock-index Coulomb potential, stock-index Coulomb potential and stock-stock Coulomb potential, along with a stock Coulomb correlation term in the Hamiltonian operator (no exchange correlation potential resulted from the absence of antisymmetric requirement in the stock market, except for the classical Coulomb interaction and Coulomb correlation potential). Finally, the conceive and feasibility of solving the stock molecular system by density functional method (DFT) are proposed together with proof. The eigen energy obtained eventually is the ground-state energy of the stock molecular system (the ground state refers to the equilibrium market). At last, we performed experiments in CSI 300 index maket and derived informations by Self-Consistent Field method (SCF). Next, we begin with the establishment of the stock molecular system proposed in this paper, accompanied by some underlying assumptions.

## 2. THE STOCK MOLECULAR SYSTEM

In quantum finance, the wave function is a function that depicts the state of a financial object. Different from previous researches, we define the Schrödinger equation of stock molecular system as follows

$$\hat{H}\Phi(r_1, r_2, \cdots, r_n, R_1, R_2, \cdots, R_N) = E\Phi(r_1, r_2, \cdots, r_n, R_1, R_2, \cdots, R_N) \tag{1}$$

Where $r_i$ refers to the return of the *i-th* stock, $i = 1, 2, \cdots, n$; $R_j$ refers to the return of the *j-th* stock index, $j = 1, 2, \cdots, N$. In Chinese stock market (except for Initial Public Offerings, the STAR Market and the GE Market, etc.), the return has the following restriction

$$\begin{cases} r_i \epsilon\ [-10\%, 10\%], i = 1, 2, \cdots, n \\ R_j \epsilon\ [-10\%, 10\%], j = 1, 2, \cdots, N \end{cases}$$

In addition, *E* is the eigenvalue which represents the ground-state energy of the stock molecular system, can be expressed by Dirac notation as $E = \langle \Phi | \hat{H} | \Phi \rangle$; $\hat{H}$ is the Hamiltonian of the stock molecular system, defined as follows

$$\hat{H} = \sum_i^n -\frac{\hbar^2}{2m_i}\nabla_i^2 + \sum_j^N -\frac{\hbar^2}{2m_j}\nabla_j^2 + \sum_i v(r_i) + \tilde{V}(R) + V(r) \tag{2}$$

where

$$\begin{cases} \nabla_i^2 = \dfrac{\partial^2}{\partial r_i^2} \\ \nabla_j^2 = \dfrac{\partial^2}{\partial R_j^2} \\ v(r_i) = \sum_j D(R_j, r_i) \\ \tilde{V}(R) = \sum_{\substack{j,z \\ m_j > m_z}} D(R_j, R_z) \\ V(r) = \sum_{\substack{i,z \\ m_i > m_z}} D(r_i, r_z) \end{cases} \tag{3}$$

$\hbar$ is the reduced Planck constant; $m_i$ is the market value of the *i-th* stock (stock index). Herein the market value is viewed as the particle mass in line with the definition of motion inertia (the larger the market value of the stock is, the harder it is to change the original direction of movement); $\nabla^2$ is the Laplace operator, $v(r_i)$ is the potential energy term of interaction between the stock and stock index, $V(r)$ is the potential energy term of interactions between stock and stock, $\tilde{V}(R)$ is the potential energy term of interactions between stock index and stock index. As the stock indexes generally rise or fall mildly whose market value is far larger than that of any single constituent stock, the Born-Oppenheimer approximation can be imported into the stock molecular system, as a result $\hat{H}$ is converted to

$$\hat{H} = \sum_i^n -\frac{\hbar^2}{2m_i}\nabla_i^2 + \sum_i v(r_i) + \tilde{V}(R) + V(r) \tag{4}$$

According to statistics, we assume that the regression relationship between variables *x* and *y* is as follows

$$\begin{cases} \hat{y} = a + kx \\ k = \rho_{xy}\dfrac{\sigma(y)}{\sigma(x)} \end{cases} \tag{5}$$

wherein

$$\begin{cases} a = E(y) - kE(x) \\ k = \dfrac{\rho_{xy}\,\sigma(y)}{\sigma(x)} \\ \rho_{xy} = \dfrac{Cov(x,y)}{\sigma(x)\sigma(y)} \end{cases} \tag{6}$$

where *a* is the intercept, *E* refers to the expectation of the stock (stock index) return; *k* is the slope, $\rho$ is the correlation coefficient, satisfying $-1 \leq \rho \leq 1$, *Cov* refers to the covariance of returns between stocks (stock indexes), $\sigma$ is the standard deviation of the stock (stock index) return. It is assumed that only stocks (stock index) with a larger market value are capable of attracting or repelling stocks (stock index) with a smaller market value. As illustrated in equation (5), for one standard deviation $\sigma(x)$ increase in the variable $x$, the variable $y$ would increase by $\rho_{xy}$ standard deviations $\sigma(y)$. Hence we define $D(x,y)$ as

$$D(x, y) = \begin{cases} a + kx - y, & \rho_{xy} \neq 0 \\ 0, & \rho_{xy} \in \{0,1\} \end{cases} \tag{7}$$

In the bargain, the real multi-electrons wave function $\Phi$ could be linearly expanded to the sum of a sequence of basis functions standing for each stock market state

$$\Phi = \sum_m c_m \varphi_m \tag{8}$$

where $\varphi_m$ are a series of basis functions signifying each possible return distribution in stock market which are defined in Hilbert space $H$, $c_m$ is the corresponding coefficient, satisfying the following equation

$$<\varphi_i|\varphi_j> = \int \cdots \iiint_{-10\%}^{10\%} \varphi_i^*(r_1, \cdots, r_n, R_1, \cdots, R_N) \varphi_j(r_1, \cdots, r_n, R_1, \cdots, R_N) \, dr_1 \cdots dr_n dR_1 \cdots dR_N$$

$$= \begin{cases} 1, & i = j \\ 0, & i \neq j \end{cases} \tag{9}$$

Contrary to the identical particle hypothesis of quantum mechanics, we defined $n(x)$ as

$$n(x) = \int \cdots \iiint_{-10\%}^{10\%} |\Phi(x, \cdots, r_n, R_1, \cdots, R_N)|^2 dr_2 \cdots dr_n dR_1 \cdots dR_N$$

$$+ \int \cdots \iiint_{-10\%}^{10\%} |\Phi(r_1, x, \cdots, r_n, R_1, \cdots, R_N)|^2 dr_1 dr_3 \cdots dr_n dR_1 \cdots dR_N$$

$$+$$
$$\vdots$$

$$+ \int \cdots \iiint_{-10\%}^{10\%} |\Phi(r_1, r_2, \cdots, r_n, R_1, \cdots, x)|^2 dr_1 \cdots dr_n dR_1 \cdots dR_{N-1} \tag{10}$$

which is the probability density of the stock (stock index) return appears at $x$.

## 3. THE SINGLE STOCK (STOCK INDEX) SCHRÖDINGER EQUATION

We derive the single stock (stock index) Schrödinger equation as follows

$$\hat{H}_i \psi_i(x) = \varepsilon_i \psi_i(x), i = 1,2, \cdots, n + N \tag{11}$$

where $\psi_i(x)$ is the state wave function of the *i-th* stock (stock index), $\varepsilon_i$ is the corresponding energy; $\hat{H}_i$ could be expressed as follows

$$\begin{cases} \hat{H}_i = -\dfrac{\hbar^2}{2m_i}\nabla_i^2 + v(r_i) + \widetilde{V}(R) + V(r_i)\,, i = 1,2,\cdots,n \\ \hat{H}_j = -\dfrac{\hbar^2}{2m_j}\nabla_j^2 + \widetilde{V}(R_j)\,, j = n+1, n+2,\cdots, N \end{cases} \quad (12)$$

Inside the formula above,

$$\begin{cases} V(r_i) = \displaystyle\sum_{z=1}^{n} \int n_z(r_z)\delta(r_z, r_i)\, dr_z + \zeta(r_i) \\ \widetilde{V}(R_j) = \displaystyle\sum_{z=n+1}^{N} \int n_z(r_z)\delta(R_z, R_j)\, dR_z + \xi(R_j) \end{cases} \quad (13)$$

where $\zeta(r_i)$ and $\xi(R_j)$ are the Coulomb correlation between stocks (stock indexes), on account of the fact that the interaction between stocks (stock indexes) would reduce the probability of that the stock (stock index) return appear at a certain value. Wherein we defined $\delta(x, y)$ as

$$\delta(x, y) = \begin{cases} D(x, y)\,, m_x > m_y \\ 0\,, m_x \le m_y \end{cases} \quad (14)$$

here the Born density $n_i(x)$ of a single stock (stock index) is defined as $n_i(x) = \psi_i^*(x)\psi_i(x)$, $i = 1, 2, \cdots, n+N$; the total Born density $n(x)$ of the whole system is $n(x) = \sum_{i=1}^{n+N}\psi_i^*(x)\psi_i(x)$. Naturally while taking the motion of stock indexes into account, we should get rid of Born-Oppenheimer approximation for the second formula in equation (12).

## 4. THE POPULARIZATION OF DENSITY FUNCTIONAL THEORY

Imatating Hohenberg and Kohn's proof [18] by *reductio ad absurdum*, the proof here is tailored for stocks, and that for stock indexes is conducted similarly. It is assumed that there exist two different external potentials $v_1(r_i)$, $\widetilde{V}_1(R)$, $v_2(r_i)$ and $\widetilde{V}_2(R)$, corresponding to the same density $\{n_i(r_i)\}$, $i = 1, 2, \cdots, n$, the two different eigen energies $E_1$, $E_2$ and the two different Hamiltonians $\hat{H}_1$, $\hat{H}_2$ of the system, $\hat{H}_1$ and $\hat{H}_2$ are manifested below

$$\begin{cases} \hat{H}_1 = \displaystyle\sum_{i}^{n} -\dfrac{\hbar^2}{2m_i}\nabla_i^2 + \sum_i v_1(r_i) + \widetilde{V}_1(R) + V(r) \\ \hat{H}_2 = \displaystyle\sum_{i}^{n} -\dfrac{\hbar^2}{2m_i}\nabla_i^2 + \sum_i v_2(r_i) + \widetilde{V}_2(R) + V(r) \end{cases} \quad (15)$$

now clearly $\Phi_1 \ne \Phi_2$ unless $\hat{H}_1 - \hat{H}_2$ equal to a constant. In the light of the variational principle, we can deduce that

$$E_1 = \langle\Phi_1|\hat{H}_1|\Phi_1\rangle < \langle\Phi_2|\hat{H}_1|\Phi_2\rangle = \langle\Phi_2|\hat{H}_2 + \textstyle\sum_i v_1(r_i) + \widetilde{V}_1(R) - \sum_i v_2(r_i) - \widetilde{V}_2(R)|\Phi_2\rangle$$

Owing to

$$\langle\Phi|\sum_i v(r_i)|\Phi\rangle = \int \cdots \iiint \Phi^* \sum_i v(r_i) \Phi d r_1 dr_2 \cdots dr_{i-1} dr_{i+1} \cdots dr_n dr_i$$

$$= \sum_i \int \cdots \iiint |\Phi|^2 v(r_i) dr_1 dr_2 \cdots dr_{i-1} dr_{i+1} \cdots dr_n dr_i$$

$$= \sum_i \int n_i(r_i) v(r_i) dr_i \tag{16}$$

it could be infered that

$$E_1 < E_2 + \sum_i \int n_i(r_i)[v_1(r_i) - v_2(r_i)]dr_i + \widetilde{V}_1(R) - \widetilde{V}_2(R) \tag{17}$$

Similarly, we have

$$E_2 < E_1 + \sum_i \int n_i(r_i)[v_2(r_i) - v_1(r_i)]dr_i + \widetilde{V}_2(R) - \widetilde{V}_1(R) \tag{18}$$

Adding up the equation (17) and (18), it could be concluded that with contradiction

$$E_1 + E_2 < E_2 + E_1 \tag{19}$$

Therefore, it could be found that the ground-state energy $E$, wave function $\Phi$ and any other information of the stock molecular system are uniquely determined by the ground-state density $\{n_i(r_i)\}$, that is, they are all the functionals of the ground-state density $\{n_i(r_i)\}$. The ground-state energy $E$ of the stock molecular system solved by DFT is enunciated as follows

$$E_{DFT} = T[\{n_i(r_i)\}] + \sum_i \langle\psi_i|v(r_i) + \widetilde{V}(R)|\psi_i\rangle + \sum_{z,i} \int\int n_i(r_i)n_z(r_z)\delta(r_z, r_i) \, dr_z dr_i + E_c[\{n_i(r_i)\}] \tag{20}$$

where $T[\{n_i(r_i)\}]$ is kinetic energy functional and $E_c[\{n_i(r_i)\}]$ is correlation functional. The formula above purport that as long as the exact functional is found, various properties of the stock molecular system shall be solved.

## 5. EXPERIMENT RESULTS

Experiments were conducted in CSI 300 index (established by China Securities Index Co.,Ltd) system which could be considered as a monatomic system for the reason that there merely exist one stock index with its constituent stocks, solved by SCF method of which the schematic process is illuminated in Figure 1. The dataset from 2023-3-31 to 2023-9-30 sourced from BaoStock and East Money website was chose to be analysed upon three simplifications: (1) The Coulomb correlation is neglected in virtue of that it is an unknown functional to everyone temporarily. (2) Each stock (stock index) is qualified to attract or repel other stocks (stock indexes). (3) The market capitalization in Hamiltonian was normalized by variant Sigmoid function as follows

$$S(x) = \frac{1}{1000[1 + exp(-0.1x)]} \tag{21}$$

taking into consideration the variance of which is too enormous to warrant the outcome authenticity in line with actuality.

At last we end up with the Born density $n(x)$ and ground-state energy of the whole stock molecular system. The informations about the top six constituent stocks in regard to market capitalization are demonstrated in Figure 2 and Table 1.

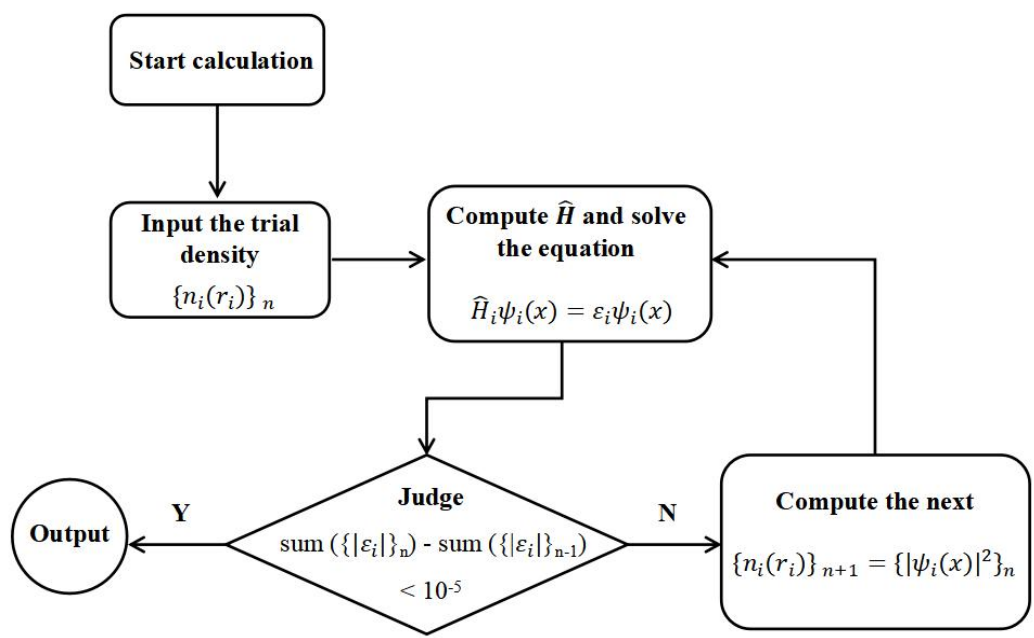

Figure 1. The schematic process of SCF method.

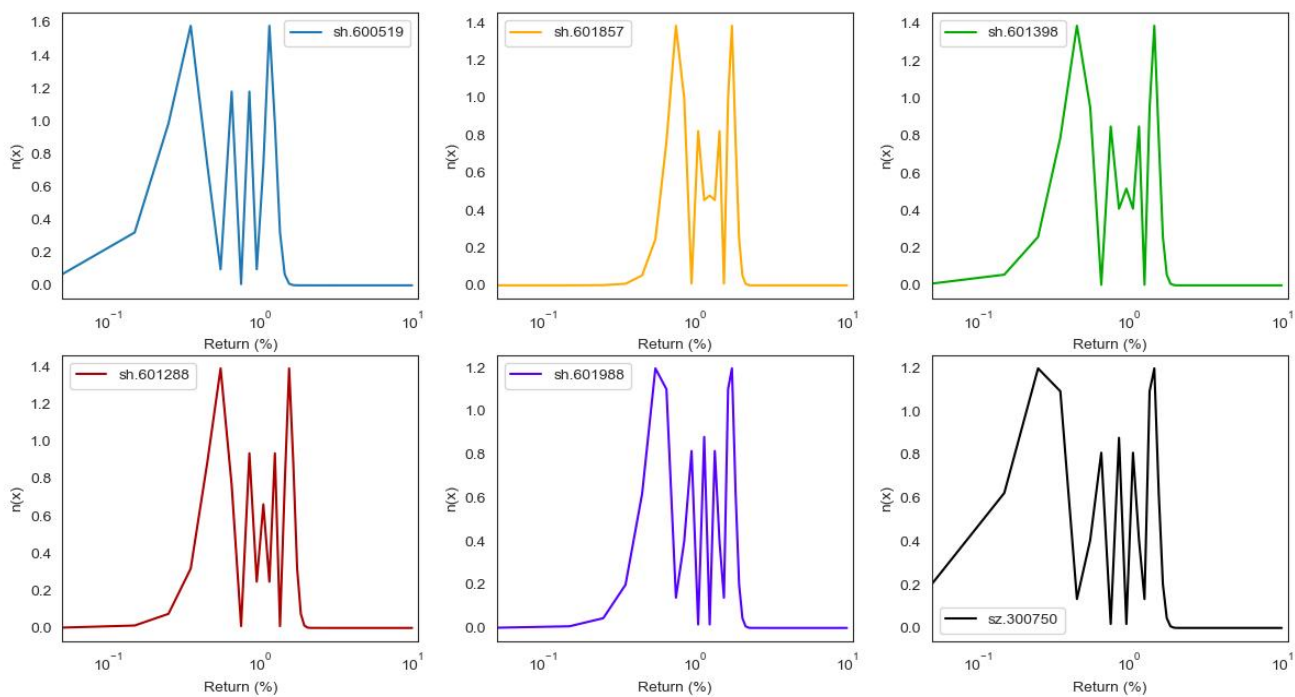

Figure 2. The Born density n(x) of the top six constituent stocks in regard to market capitalization with respect to stock return which is logarithmic coordinate.

Figure 2 is of semi-logarithm coordinates, in which the local maximum points and minimum points of the curves betoken the stock return would occur at one awfully probably and impossibly in the future. Additionally, we could leard that the distribution of returns differ from each other among this six constituent stocks.

Table 1. The ground-state energy of the top six constituent stocks in regard to market capitalization.

| Stock | Energy |
|---|---|
| Sh.600519 | 5.0964 |
| Sh.601857 | -1.8137 |
| Sh.601398 | -0.5281 |
| Sh.601288 | 10.0478 |
| Sh.601988 | -2.7935 |
| Sz.300750 | 9.9856 |

The Born density $n(x)$ symbolise the probability of the stock return being $x$, and the ground-state energy stand for the tendency of the stock return being about to rise or fall in the future while positive or negative.

# 6. CONCLUSION

In this paper, to begin with the construction of the stock molecular system under the framework of quantum mechanics, the entire stock market is treated as a large molecule, in which all stock indexes are treated as nuclei in the molecule, and all constituent stocks assigned to their stock indexes are treated as electrons. We constructed the Hamiltonian of the stock molecular system with bringing Born-Oppenheimer approximation in. The description of interaction among stocks (stock indexes) rely on the interactive potential terms which are constructed by taking advantage of the regression statistics theory (in fact, other statistical relationships are exploitable). Ultimately, the Schrödinger equation of a single stock (stock index) and its corresponding Hamiltonian are elucidated. The Born density of the stock molecular system is defined under the non-identical particles hypothesis. Furthermore, it is proved by *reductio ad absurdum* that various properties of the stock molecular system are still uniquely determined by the density $\{n_i(r_i)\}$, which popularizes the Density Functional Theory to the research of financial market. Based on the final experiments carried out in CSI 300 index system, the methodology presented in this paper can indeed be applied to acquire the informations of the whole stock market. Our article provides the new methodology and ideology for the study of finance and economy.

**REFERENCES**

[1] R.N. Mantegna, H.E. Stanley, Scaling behaviour in the dynamics of an economic index, Nature 376 (6535) (1995) 46-49.
[2] R.N. Mantegna, H.E. Stanley, N.A. Chriss, An introduction to econophysics: Correlations and complexity in finance, Phys. Today 53 (12) (2000) 70.
[3] S.N. Durlauf, Statistical mechanics approaches to socioeconomic behavior, in: W.B. Arthur, S.N. Durlauf, D.A. Lane (Eds.), The Economy as an Evolving Complex System II, CRC Press, Boca Raton, Florida, 2018, pp. 81–104.
[4] K. Ilinski, Physics of Finance: Gauge Modeling in Non-equilibrium Pricing, John Wiley & Sons, New York, 2001,
[5] Dash, J. (1988). Path Integrals and Options, Part I, CNRS Preprint CPT-88/PE.2206.
[6] R.P. Feynman, A.R. Hibbs, Quantum Mechanics and Path Integrals, McGraw-Hill, New York, 1965.

[7] Esmailzadeh, R. (1995). Path-Dependent Options, Morgan Stanley Report.
[8] B.E. Baaquie, A path integral approach to option pricing with stochastic volatility: Some exact results, J. Phys. I 7 (12) (1997) 1733-1753.
[9] F. Black, M. Scholes, The pricing of options and corporate liabilities, J. Polit. Econ. 81 (3) (1973) 637-654.
[10] E.E. Haven, A discussion on embedding the Black‐Scholes option pricing model in a quantum physics setting, Physica A 304 (3-4) (2002) 507-524.
[11] M. Schaden, Quantum finance, Physica A 316 (1-4) (2002) 511-538.
[12] C. Zhang, L. Huang, A quantum model for the stock market, Physica A 389 (24) (2010) 5769-5775.
[13] L.A. Cotfas, A finite-dimensional quantum model for the stock market, Physica A 392 (2) (2013) 371-380.
[14] C. Ye, J.P. Huang, Non-classical oscillator model for persistent fluctuations in stock markets, Physica A 387 (5-6) (2008) 1255-1263.
[15] C. Chaiboonsri, S. Wannapan, Applying quantum mechanics for extreme Value prediction of VaR and ES in the ASEAN stock exchange, Economies 9 (1) (2021) 13.
[16] K. Ahn, M.Y. Choi, B. Dai, S. Sohn, B. Yang, Modeling stock return distributions with a quantum harmonic oscillator, EPL 120 (3) (2017) 38003.
[17] T. Gao, Y. Chen, A quantum anharmonic oscillator model for the stock market, Physica A 468 (2017) 307-314.
[18] P. Hohenberg, W. Kohn, Inhomogeneous Electron Gas, Phys. Rev. 136 (3B) (1964) B864-B871.